\documentclass{aa}
\usepackage{graphicx}
\usepackage{txfonts}
\begin{document}
   \title{Technetium and the third dredge up in AGB stars}

   \subtitle{I. Field stars\thanks{Partly based on observations collected
   at the European Southern Observatory, Paranal, Chile (ESO-Programme 65.L-0317(A))}}

   \author{T. Lebzelter\thanks{Visiting Astronomer, Kitt Peak
   National Observatory, NOAO, which is operated by the Association
   of Universities for Research in Astronomy, Inc. (AURA) under cooperative
   agreement with the National Science Foundation.}
          \inst{1}
          \and
          J. Hron\inst{1}
          }

   \offprints{T. Lebzelter}

   \institute{Institute for Astronomy (IfA), University of Vienna,
              T\"urkenschanzstrasse 17, A-1180 Vienna\\
              \email{lebzelter@astro.univie.ac.at}
             }

   \date{Received ; accepted }

   \abstract{We searched for Tc in a sample of long period variables selected
   by stellar luminosity derived from Hipparcos parallaxes. Tc, as an unstable
   s-process element, is a good indicator for the evolutionary status of
   stars on the asymptotic giant branch (AGB). In this paper we study the
   occurrence of Tc as a function of luminosity to provide constraints on the minimum luminosity
   for the third dredge up as estimated from recent stellar evolution models.

   A large number of AGB stars above the estimated theoretical 
   limit for the third dredge
   up are found not to show Tc.
   We confirm previous findings that only a small fraction of the semiregular
   variables show Tc lines in their spectra. Contrary to earlier results by 
   Little et al.\,(1987) we find also a significant number of Miras without Tc.

   The presence and absence of Tc is discussed in relation to the mass distribution of AGB stars.
   We find that a large fraction of the stars of our sample must have current masses
   of less than 1.5\,$M_{\sun}$. Combining our findings with stellar evolution scenarios
   we conclude that the fraction of time a star is observed as a SRV or a Mira is
   dependent on its mass.
   \keywords{stars: late-type -- stars: AGB and post-AGB -- stars: evolution                
               }
   }

   \maketitle
%
%________________________________________________________________

\section{Introduction}
The Asymptotic Giant Branch (AGB) phase is an important step in the final evolution for the 
majority of stars. In the most luminous part of the AGB the behavior of a star is characterized
by the so called Thermal Pulses (TP), thermal instabilities of the He shell accompanied by changes in luminosity, temperature, period and
internal structure (see e.g.\ Busso et al.~\cite{bugawa99} for a review).
Between the repeated events of explosive He-burning heavy elements can be produced via the s-process in the region between the hydrogen and the helium shells. The freshly produced material is then brought to the stellar
surface by the convective envelope that temporarily extends to these very deep layers (3$^{rd}$
dredge up; 3DUP). This dredge up is responsible for changing the elemental abundances of the 
stellar atmosphere from oxygen rich into carbon rich. 

During the last years considerable progress has been made with regard to models for the 3DUP and nucleosynthesis on the thermally pulsing AGB (TP-AGB; Busso et al.~\cite{bugawa99}, Lugaro et al.~\cite{lug03} and references therein). The different evolution models agree qualitatively in the sense that the 3DUP is more efficient for more massive convective envelopes (e.g.\ Straniero et al.~\cite{scl97}) and for lower metallicities (e.g.\ Busso et al.~\cite{bus01}). However, the quantitative results are still model dependent (Lattanzio \cite{lattanzio02}, Lugaro et al.~\cite{lug03}) and grids of new models covering a wider range of stellar parameters are scarce. In spite of the observational and theoretical 
uncertainties, the observed s-element abundances of AGB-stars agree with the model predictions and thus support the metallicity dependence of the 3DUP (Busso et al.~\cite{bus01}, Abia et al.~\cite{abia02}). A first attempt to directly check the conditions for the onset of 3DUP observationally has been made by Lebzelter \& Hron (\cite{lh99}). Important constraints on the minimum (core) mass (and hence luminosity) for 3DUP and its efficiency also come from the observed luminosity function of carbon stars
in the LMC and synthetic stellar evolution calculations.

Among the elements produced during the TP-AGB is $^{99}$Tc, a radioactive element with
a half life time of only 2.10$^5$ years. This fact makes it to a reliable indicator of the 3DUP,
because due to the short life time any Tc we see in a star has been produced during its previous evolution on the TP-AGB. Technetium should be detectable at the surface after only a few thermal pulses (Goriely \& Mowlavi \cite{gm00}). It should be noted at this point that the absence of Tc does not necessarily mean the absence of TPs but rather the absence of 3DUP for several TPs. This could be caused by a too low initial mass on the TP-AGB or by a too high mass loss rate at the end of the AGB-evolution. We will come back to this point later on.

Many long period variables (Miras, semiregular variables, irregular variables), which
are thought to be on the AGB, have been searched for Tc lines in their spectra (Little et al.\,\cite{llmb87},
Lebzelter \& Hron \cite{lh99} and references therein) to check for a relation between variability and dredge up. Miras with periods
of more than 300 days were found to show Tc in their spectra, while most semiregular variables (SRVs) do
not show Tc. Lebzelter \& Hron (\cite{lh99}) argued that the small fraction of SRVs with Tc are due to a
contribution of high mass objects to this class of variables. The majority of the SRVs are
low mass stars that have not yet reached the minimum core mass (or equivalently the necessary luminosity)
predicted for the 3DUP. The results of Lebzelter \& Hron (\cite{lh99}) were a first indication that the
luminosity limit derived from LMC stars and stellar evolution models actually also applies
to galactic AGB stars.

Among the stars with reliable Hipparcos parallaxes are still several AGB stars which are close to
or brighter than the theoretical 3DUP luminosity, have enough other observational data to reliably
estimate their bolometric magnitudes, but have not yet been searched for Tc. In the light of the small number of stars with good
Hipparcos parallaxes and results on the presence of Tc a high demand exists to fill this observational
gap. In this paper we will present measurements on the occurrence of Tc on an widely extended sample
of M-type AGB stars with Hipparcos parallaxes.

Another group of AGB stars that has hardly been investigated for Tc lines in their spectra are
Miras with periods above 400 days. These variables are of special interest for the question of
the 3DUP limit as observed period luminosity relations for Miras (e.g.~Alvarez \& Mennessier \cite{am97})
suggest that these stars should be above the luminosity limit and should therefore all show
Tc in their spectra. In the present paper we also give new observational results on
these long period Miras.

\section{The sample}
\subsection{Observations}
We selected a sample of AGB stars with measured parallaxes in the Hipparcos
Catalogue (ESA \cite{esa97}). Only those stars were used for which the parallaxes
followed the criteria listed in Table \ref{t:hippcrit}. These criteria were chosen
to exclude all stars with senseless parallaxes (trigonometric parallax $<$0) or 
bad astrometric data due to modelling errors (goodness of fit). 
The maximum ratio of the error of the parallax over the parallax was
selected to allow correction of the Lutz-Kelker bias (see below). Primary targets were those stars
that have not been searched for Tc in the past, and objects without a definite
decision on their Tc content. The selection of Miras with periods longer than 400 days
was based on the General Catalogue of Variable Stars (Kholopov et al.\,\cite{GCVS}, GCVS).

\begin{table}
\caption{Selection criteria for AGB stars with good parallaxes from
the Hipparcos catalogue (ESA 1997). The third column gives the corresponding
sources as coded in the Hipparcos catalogue.}  \label{t:hippcrit}
%\begin{flushleft}
\begin{tabular}{lcl}
\hline 
\noalign{\smallskip}
Goodness of fit & $\lid$3.14 & H30\\
Trigonometric parallax & $>$0 & H11\\
$\sigma_{\pi}$/$\pi$ & $\lid$0.5 & H11, H16\\
\noalign{\smallskip}
\hline 
\end{tabular}
%\end{flushleft}
\end{table}

\begin{table*}
\caption{Observing log. Given are the name of the variable (column 1), its variability
type (column 2), its spectral type (column 3) and its period (column 4), respectively.
All these data are taken from the General Catalogue of Variable Stars (GCVS, Kholopov
et al.\,\cite{GCVS}. Column 5 lists whether or not Technetium could be detected
in the obtained spectra. The last column gives the instrument used: KPNO - Coude Feed at
KPNO, AAT - UCLES at the AAT, UVES - UVES at the VLT.}  \label{t:obslog}
%\begin{flushleft}
\begin{tabular}{lllrcl}
GCVS name    & Var.type    & Spectral type     & Period [d] & Tc? & Instrument\\
\hline 
\noalign{\smallskip}
\object{AqlRR}        &    M        &    M6E-M9         & 394.78 &    no          &   KPNO\\
\object{AqlV844}      &    SRa      &    M5-M7EP        & 369.00 &    no          &   KPNO\\
\object{AqrR}         &    M        &    M5E-M8.5E      & 386.96 &    yes         &   KPNO\\
\object{AriT}         &    SRa      &    M6E-M8E        & 316.60 &   no          &   KPNO\\
\object{AurNO}        &    Lc       &    M2SIAB         & &    no          &   KPNO\\
\object{AurRU}        &    M        &    M7E-M9E        & 466.47 &    yes         &   KPNO    \\
\object{BooRX}        &    SRb      &    M6.5E-M8IIIE   & 340.00 &    no          &   KPNO\\
\object{BooV}         &    SRa      &    M6E            & 258.01 &    no          &   KPNO\\
\object{CamRV}        &    SRb      &    M4II-III-M6    & 101.00 &    no          &   KPNO\\
\object{CapAG}        &    SRb      &    M3III          & 25.00 &    no          &   KPNO,AAT\\
\object{CapT}         &    M        &    M2E-M8.2       & 269.28 &    no          &   KPNO\\
\object{CasAA}        &    Lb       &    M6III          &  &    no          &   KPNO\\
\object{CasVY}        &    SRb      &    M6-M7          & 100.00 &    no          &   KPNO\\
\object{CetAG}        &    SRb      &    M3             & 90.00 &    no          &   KPNO\\
\object{CetAM}        &    SRb      &    M5III          & 70.00 &    no          &   KPNO\\
\object{CetAT}        &    SRb      &    M5             & 60.00 &    no          &   KPNO,AAT    \\
\object{CetU}         &    M        &    M2E-M6E        & 234.76 &    no          &   UVES    \\
\object{CncBL}        &    Lb       &    M3III          & &    no          &   KPNO\\
\object{CncBO}        &    Lb:      &    M3III          & &    no          &   KPNO\\
\object{CncBP}        &    SRb      &    M3III          & 40.00 &    no          &   KPNO    \\
\object{CncX}         &    SRb      &    C5,4(N3)       & 195.00 &    no          &   KPNO\\
\object{CrBRS}        &    SRa      &    M7             & 332.20 &    no          &   KPNO\\
\object{CrBY}         &    SRb      &    M8             & 300.00 &    no          &   KPNO\\
\object{CygW}         &    SRb      &    M4E-M6E        & 131.10 &    yes         &   KPNO\\
\object{DelCT}        &    Lb       &    M7             & &    no          &   KPNO,AAT\\
\object{DelU}         &    SRb      &    M5II-III       & 110.00 &    yes         &   KPNO\\
\object{DraCQ}        &    Lb:      &    M3IIIA         & &    no          &   KPNO\\
\object{DraWZ}        &    SRa      &    M6E            & 401.70 &    yes         &   KPNO    \\
\object{EriCY}        &    SRb      &    M4III          & 25.00 &    no          &   KPNO,AAT\\
\object{EriDQ}        &    SRb      &    M4III          & 30.00 &    no          &   KPNO\\
\object{EriUU}        &    SRb      &    M7             & 340.00 &    no          &   KPNO\\
\object{EriW}         &    M        &    M7E-M9         & 376.63 &    possible    &   UVES    \\
\object{EriZ}         &    SRb      &    M4III          & 80.00 &    no          &   KPNO\\
\object{GemNP}        &    Lb:      &    M1.5           & &     no          &   KPNO\\
\object{GemNZ}        &    SR       &    M3II-IIIS      & &     no          &   KPNO\\
\object{GemSW}        &    SRa      &    M5III          & 680.00 &    no          &   KPNO\\
\object{GruS}         &    M        &    M5E-M8IIIE     & 401.51 &    yes         &   UVES    \\
\object{Herg}         &    SRb      &    M6III          & 89.20 &    no          &   KPNO\\
\object{HerRU}        &    M        &    M6E-M9         & 484.83 &    yes         &   KPNO\\
\object{HerST}        &    SRb      &    M6-7IIIAS      & 148.00 &    yes         &   KPNO\\
\object{HerU}         &    M        &    M6.5E-M9.5E    & 406.10 &    possible    &   KPNO\\
\object{HerV337}      &    SRb      &    M8             & 280.00 &    no          &   KPNO\\
\object{HorR}         &    M        &    M5E-M8EII-III  & 407.60 &    yes         &   UVES    \\
\object{HorT}         &    M        &    M5IIE          & 217.60 &    no          &   UVES    \\
\object{HorTW}        &    SRb      &    C7,2(N0)       & 158.00 &    yes         &   AAT,UVES\\
\object{HyaIN}        &    SRb      &    M4III          & 65.00 &    no          &   KPNO\\
\object{HyaR}         &    M        &    M6E-M9ES       & 388.87 &    yes         &   UVES    \\
\object{HyaRU}        &    M        &    M6E-M8.8E      & 331.50 &    no          &   UVES    \\
\object{LacRX}        &    SRb      &    M7.5SE         & 650.00 &    yes         &   KPNO\\
\object{LacTV}        &    Lb       &    C4,5(N3)       & &    no          &   KPNO\\
\object{LeoDE}        &    SRb:     &    M2IIIABS       & &     no          &   KPNO    \\
\object{LeoDF}        &    SRb      &    M4III          & 70.00 &    no          &   KPNO\\
\object{LeoVY}        &    Lb:      &    M5.5III        & &    no          &   KPNO\\
\object{LibY}         &    M        &    M5E-M8.2E      & 275.70 &    no          &   UVES    \\
\object{LMiRX}        &    SRb      &    M4IIIA         & 150.00 &    no          &   KPNO    \\
\object{LynRT}        &    M        &    M6E            & 394.60 &    yes         &   KPNO\\
\object{LynU}         &    M        &    M7E-M9.5:E     & 433.60 &    no          &   KPNO\\
\noalign{\smallskip}
\hline
\end{tabular}
\end{table*}

\setcounter{table}{1}

\begin{table*}
\caption{Continued.}
%\begin{flushleft}
\begin{tabular}{lllrcl}
GCVS name    & Var.type    & Spectral type     & Period [d] & Tc? & Instrument\\
\hline 
\noalign{\smallskip}
\object{LynUX}        &    SRb:     &    M6III          & &    no          &   KPNO\\
\object{LynUY}        &    Lb:      &    M3IIIAB        & &    no          &   KPNO\\
\object{LyrV398}      &    Lb       &    M6             & &    no          &   KPNO    \\
\object{MicT}         &    SRb      &    M6E            & 347.00 &    no          &   KPNO,AAT\\
\object{MicU}         &    M        &    M5E-M7E        & 334.29 &    no          &   UVES    \\
\object{MonSY}        &    M        &    M6E-M9         & 422.17 &    no          &   KPNO\\
\object{MonX}         &    SRa      &    M1EIII-M6EP    & 155.80 &    no          &   KPNO\\
\object{OriW}         &    SRb      &    C5,4(N5)       & 212.00 &    no          &   KPNO\\
\object{PavNU}        &    SRb      &    M6III          & 60.00 &    no          &   AAT    \\
\object{PavS}         &    SRa      &    M7IIE-M8III    & 380.86 &    possible    &   AAT\\
\object{PegBD}        &    SRb      &    M6-M8          & 78.00 &    no          &   KPNO\\
\object{PegGO}        &    Lb       &    M4             & &    no          &   KPNO\\
\object{PegS}         &    M        &    M5E-M8.5E      & 319.22 &    no          &   KPNO    \\
\object{PegTW}        &    SRb      &    M6-M8          & 929.30 &    no          &   KPNO\\
\object{PegUW}        &    SRb      &    M5-M7          & 106.00 &    yes         &   KPNO\\
\object{PegW}         &    M        &    M6E-M8E        & 345.50 &    no          &   KPNO\\
\object{PerUZ}        &    SRb      &    M5II-III       & 927.00 &    no          &   KPNO\\
\object{PscTW}        &    Lb       &    M8             & &    no          &   KPNO\\
\object{PscTX}        &    Lb       &    C7,2(N0)       & &    yes         &   KPNO\\
\object{SclY}         &    SRb      &    M4             & &    no          &   UVES    \\
\object{SerDX}        &    SRa      &    M5-M8          & 360.00 &    no          &   KPNO\\
\object{SerY}         &    SRa      &    M5IIIE         & 432.70 &    no          &   KPNO\\
\object{SgeT}         &    SRb      &    M4-M6.5        & 165.50 &    yes         &   KPNO\\
\object{SgrSU}        &    SRb      &    M6III          & 60.00 &    no          &   KPNO,AAT\\
\object{TucT}         &    M        &    M3IIE-M6IIE    & 250.30 &    no          &   UVES    \\
\object{UMaCG}        &    Lb       &    M4IIIA         & &    no          &   KPNO\\
\object{UMaCS}        &    Lb:      &    M3IIIAB        & &    no          &   KPNO\\
\object{UMaTV}       &    SRb      &    M5III          & 42.00 &    no          &   KPNO\\
\object{UMaVW}        &    SR       &    M2             & 610.00 &    no          &   KPNO\\
\object{VirER}	       &    SRb	     &    M4III          & 55.00 &    no	         &   UVES\\
\object{VirEV}	       &    SRb	     &    M4II-III       & 120.00 &    no	         &   UVES\\
\object{VirFW}        &    SRb:     &    M3IIIAB        & 15.00 &    no          &   KPNO\\
\object{VirRS}        &    M        &    M6IIIE-M8E     & 353.95 &    no          &   UVES    \\
\object{VirS}         &    M        &    M6IIIE-M9.5E   & 375.10 &    yes         &   UVES    \\
\object{VirSW}        &    SRb      &    M7III          & 150.00 &    yes         &   KPNO    \\
\object{VulDY}        &    Lb       &    M3-M6          & &    no          &   KPNO\\
\object{VulFI}        &    Lb       &    M3             & &    no          &   KPNO    \\
\noalign{\smallskip}
\hline
\end{tabular}
\end{table*}

Observations obtained at three observatories were used. For northern objects spectra
were taken with the Coude spectrograph and the Coude Feed telescope at Kitt Peak National
Observatory (KPNO). We had two observing runs in January and July 2000, respectively. In 
both cases the spectral resolution achieved was 35000. Further observations were obtained
with Anglo-Australian Telescope (AAT) and UCLES. These observations were done in service 
mode. Resolution was 50000. Finally, some data came from an observing run in July 2000 at the ESO VLT
using UVES (R=50000). In all cases the wavelength range was chosen to include the two lines of Tc
at 4262 and 4297\,{\AA}, respectively. Data from VLT/UVES also included a third Tc line at
4236\,{\AA}. The S/N ratio achieved was typically better than 80 with a few faint stars
having S/N ratios of only $\approx$\,20. Representative spectra of stars with and without Tc taken at similar spectral resolution as in this work are shown in Fig.~1 of Lebzelter \& Hron (\cite{lh99}).

A complete list of all stars observed in the course
of this projects is given in Table\,\ref{t:obslog}. Not included are only those stars
that were used as reference objects, i.e.~for which the Tc contents has not been determined
during this investigation. A few objects have been observed at more than one observatory.
All measurements gave consistent results on the Technetium contents.

\subsection{Data from the literature}
Beside the new observations information on the occurrence of Tc for further stars 
has been taken from the literature. All measurements from the comprehensive paper
by Little et al.\,(\cite{llmb87}) have been taken into account. Little et al.~introduce
five levels in their classification on the occurrence of Tc, namely {\it yes, probable,
possible, doubtful} and {\it no}. As in Lebzelter \& Hron (\cite{lh99} we will combine {\it yes} and
{\it probable} into {\it yes} and {\it doubtful} and {\it no} into {\it no}. 

All semiregular and irregular variables presented in our previous paper (Lebzelter \& Hron \cite{lh99})
were included in our sample. Again a few stars previously classified as {\it possible} were
reobserved and the new spectra were used for the classification.
Additional information Tc-rich and Tc-poor stars came from the papers by van Eck et al.\,(\cite{EJUMP98}) 
and Abia et al.\,(\cite{abia02}). 
These literature data combined with our new measurements resulted in a sample of
345 late type (AGB) stars checked for the occurrence of Tc lines. Table\,\ref{t:sample} summarizes the
different parts of the sample.

\begin{table}
\caption{Sample description}
\label{t:sample}
\begin{tabular}{lcccc}
\hline
\noalign{\smallskip}
 & \multicolumn{3}{c}{Tc} & \\
 & yes & no & possible & total\\
\noalign{\smallskip}
\hline
\noalign{\smallskip}
total sample & 79 & 245 & 21 & 345\\
stars with abs. magnitudes & 18 & 81 & 2 & 101\\
variables with known period & 73 & 208 & 21 & 302\\
\noalign{\smallskip}
\hline
\noalign{\smallskip}
Miras & 52 & 77 & 19 & 148\\
SRa & 4 & 26 & 1 & 31\\
SRb & 16 & 106 & 1 & 148\\
SRc & 1 & 1 & 0 & 2\\
SR & 0 & 3 & 0 & 3\\
Lb & 5 & 30 & 0 & 35\\
Lc & 0 & 1 & 0 & 1\\
\noalign{\smallskip}
\hline
\end{tabular}
\end{table}

Very briefly we want to discuss here the S and C type stars. 
Stars like these are produced either by
contamination from a more massive companion in a binary system or from their own 3DUP processes. 
Small samples of stars with these spectral
types have been studied by van Eck et al.\,(\cite{EJUMP98}) and Abia et al.\,(\cite{abia02}) as well as Bergeat et al.~(\cite{bkr02}) , respectively.
The studies of van Eck et al.\ and Bergeat et al.\ indicated that most (but not all) of the stars with Tc are
found above the luminosity limit of the 3DUP. On the other hand, not all stars above the
luminosity limit showed Tc.
However, as noted by Abia et al.
(\cite{abia02}) it is extremely difficult to exclude the existence of Tc lines in the crowed spectra of
some of the C-rich stars. One S type Mira out of 15 
shows no Tc in its spectrum. 17 SRVs in our sample (including data from the papers listed
above) are carbon rich objects, in only 6 of them
Tc has been detected. Due to the difficulties in the detection and in determining the evolutionary
history of S and C stars we decided to limit our discussion to M and MS stars.
A few newly observed C-type stars are also listed but we will not include them in the further analysis.

\subsection{Bolometric magnitudes}

Bolometric magnitudes were taken from Kerschbaum \& Hron (\cite{KH96}) and Kerschbaum (\cite{kerschbaum99})
where available. In these papers a blackbody was fitted to visual, near infrared and IRAS photometry. In the case
of a notable infrared excess two blackbodies (star and dust shell) were fitted to the photometry.
Integrating over the blackbody gave the bolometric magnitude. For the remaining stars similar
blackbody fits were kindly provided by Franz Kerschbaum (private communication). Near Infrared photometry has been
taken from Bagnulo (\cite{bagnulo96}), Fouque et al.\,(\cite{fouque92}), 
Kerschbaum (\cite{kerschbaum95}), Kerschbaum \& Hron (\cite{kh94}), Kerschbaum et 
al.\,(\cite{klh96}, \cite{kll01}) and Whitelock et al.\,(\cite{wmf00}). Visual photometry
came from the GCVS.

As blackbody fits are available for all our sample objects we could directly determine the bolometric magnitudes. Nevertheless, we compared the results with bolometric magnitudes calculated from $K$ brightness and a
bolometric correction based on $J-K$ using the relation given by Montegriffo et al.\,(\cite{mfop98}) 
for $J-K\lid1.56$ and the one by Costa \& Frogel (\cite{cf96}) for $J-K>1.56$, respectively. It should be noted that the above relations are mostly based on stars not really comparable to the objects in our sample. We found that
m$_{bol}$ values based on blackbody fits are in the same range as those calculated
using bolometric corrections for small values of $J-K$, but come out at somewhat lower values
for redder stars. We will present a more detailed comparison of these two approaches in
a forthcoming paper (Kerschbaum et al.~in preparation).

To derive absolute bolometric magnitudes parallaxes were taken from the Hipparcos catalogue. 
Where available the values from the re-calibration of the Hipparcos data by Pourbaix et al.\,(\cite{platais03}) and
Knapp et al.\,(\cite{KPPJ}) were used. A correction for interstellar 
absorption was applied using the same approach as Feast et al.\,(\cite{feast82}). This correction,
however, was very small for the stars of our sample (0.01 to 0.03 mag). Finally, a correction for
Lutz Kelker bias was applied following the method outlined by Hanson (\cite{hanson79}).

Absolute K magnitudes derived from Hipparcos data have also been published earlier by Mennessier et al.\,(\cite{mmal01}).
It has to be noted that the values given by Mennessier et al.~are not simply calculated from
the parallax but use an additional (statistical) input from a classification of the star due to some 
photometric criteria (V, IRAS colors).
In general we found a quite good agreement between our values and their results, which is not
surprising as very similar sources of infrared measurements have been used. The reason for the few stars
with differing results could not be found, but as the Mennessier et al.~values are based on a statistical 
approach, such deviations for a few individual objects are not surprising.

%Sources for infrared
%photometry are given in Table \ref{allmydata} together with fundamental stellar parameters, the 
%occurrence of Tc and the absolute bolometric magnitude for each star of the sample. 

\subsection{Comparing new measurements with literature data}
Some of the stars in the observed sample have been investigated for Tc already before. We observed these
stars to check consistency in the classification and to clarify cases classified as 'possible' in the
literature. In general, the agreement between the new measurements and the literature data is very good.
Among the objects reobserved, those stars classified by LLMB as 'doubtful' or 'probable' all showed
in the new data 'no Tc' or 'Tc', respectively. This justifies the assumption made before on reducing the
five classification levels of LLMB into three.
A few objects with differing (or clarified) results exist. They are listed in Table\,\ref{t:diffres}.
In these cases the decision on the occurrence of Tc was based on our own spectra. An interesting case is
the SRV V\,Boo. LLMB had originally classified the star as 'probable'. Lebzelter \& Hron
(\cite{lh99}) found that the star has a puzzling location for a Tc rich star in the period-luminosity diagram.
Some outstanding peculiarity was suspected due to its strange light curve behavior. However, our new
spectrum of V\,Boo clearly shows no Tc lines.

\begin{table}
\caption{Comparison with literature values. Column 3 gives the reference for the literature
value: LLMB...Little et al.\,\cite{llmb87}, LH99...Lebzelter \& Hron \cite{lh99}.} \label{t:diffres}
\begin{tabular}{lccc}
\hline
\noalign{\smallskip}
GCVS name & Tc literature & reference & Tc new\\
\noalign{\smallskip}
\hline
\noalign{\smallskip}
AqlRR & possible & LLMB & no\\
BooV  & yes      & LLMB & no\\
CetU  & no       & LLMB & possible\\
HerU  & no       & LLMB & possible\\
HyaRU & possible & LLMB & no\\
MicT  & possible & LH99 & no\\
MonX  & possible & LH99 & no\\
PavS  & no       & LH99 & possible\\
VirSW & possible & LH99 & yes\\
\noalign{\smallskip}
\hline
\end{tabular}
\end{table}

\section{Results}
\subsection{The minimum luminosity for 3DUP}
Recent stellar evolution calculations for stars with solar metallicity and initial masses between 1 and 3\,M$_{\sun}$ indicate, that 3DUP occurs for core masses above 0.58--0.62\,M$_{\sun}$ (Straniero et al.~\cite{scl97}, Herwig et al.~\cite{hbd2000}). The above range in core mass mostly reflects the dependency of the core mass limit on the input physics of the calculations. The main factor driving 3DUP (besides the input physics) is the mass of the convective envelope, which has to be larger than about 0.5\,M$_{\sun}$ (see e.g.\ Straniero et al.~\cite{scl97}). Thus at solar metallicity the two different sets of calculations predict 3DUP only for stars with initial masses above 1\,M$_{\sun}$. This limit is also supported by the analysis of carbon star luminosity functions (Marigo et al.\,\cite{MGB99}) and a comparison of different model calculations by Mowlavi (\cite{Mowlavi99}). 

Applying the classical core mass luminosity relation (e.g. Marigo et al.~\cite{mgwg99}) to the above core mass limits results in a luminosity limit for 3DUP of 7000 to 9000 \,L$_{\sun}$. However, for a comparison with observations
two effects should be taken into account: (i) during the first TPs the pre-flash (quiescent) surface luminosity is significantly lower than predicted by a linear core mass luminosity relation (see Marigo et al.~\cite{mgwg99} for a discussion). This is of special relevance for low mass stars where the envelope mass is larger than the minimum only during the first few pulses.
(ii) neglecting the fast luminosity change during the TP, the surface luminosity increases between two consecutive TPs by about a factor of 2 (e.g. Straniero et al.~\cite{scl97}).

Thus instead of applying the core mass luminosity relation we have estimated luminosity limits from the luminosity evolution of the 1.50\,M$_{\sun}$ model of Straniero et al.\,(\cite{scl97}). For a core mass of  0.58\,M$_{\sun}$, the surface luminosity varies between
M$_{\rm bol}=-3\fm 9$ and $-$4$\fm 8$, for 0.62\,M$_{\sun}$ the corresponding range is from $-4\fm 2$ to $-5\fm 1$. Thus from these models we expect no stars with Tc below M$_{\rm bol}=-3\fm 9$. Above this luminosity, the fraction of Tc-rich stars should increase if the envelope mass is high enough for repeated 3DUP. However, it has to be kept in mind, that the conditions for the onset of 3DUP (and hence this luminosity limit) are still uncertain.

\subsection{Comparison with observations}
\begin{figure}
   \centering
   \includegraphics{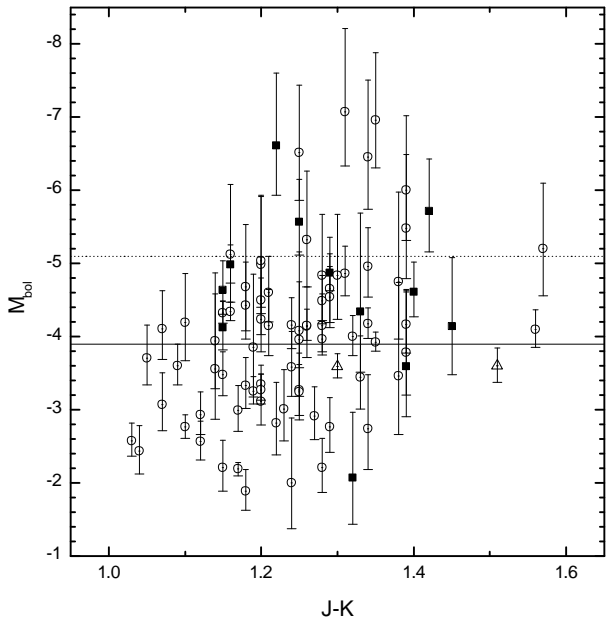}
   \caption{Colour-luminosity diagram for the M-type AGB stars with good Hipparcos
   parallax searched for Tc lines in their spectrum. Open symbols denote
   non detections, filled circles represent stars showing Tc, and filled triangles
   mark unclear cases (possible). The full line marks the approximate minimum
   luminosity of an AGB-star when third dredge up sets in. The dotted line shows
   the approximate maximum luminosity at this stage. See Sect.~3.1 for details.}
   \label{f:hrd}
    \end{figure}

In Fig.\,\ref{f:hrd} we plot a HRD of our sample of AGB stars. $J-K$ has been used
as a temperature indicator (e.g.~Bessell et al.~\cite{bsw96}, Houdashelt et al.\,\cite{HBSW00}). 
However, as noted by Bessell et al.~(\cite{bsw96}), for the cooler Miras $J-K$ is of only limited use
as a temperature indicator. Sources of near
infrared photometry have been described above. The full line in the figure marks
the approximate minimum luminosity for 3DUP as discussed above.
Error bars are derived from the error of the parallax $\sigma_{\pi}$ given in the
Hipparcos catalogue.

Taking into account the error in the parallax 
it can be seen that all but one star with Technetium are found above or very
close to the estimated theoretical third dredge up limit. The two {\it possible} cases lie slightly below the limit. It has to be taken into account that for a significant fraction of the stars no near infrared
light curves exist. This introduces a further scatter into the luminosities that has not be included in the error bars given. For those stars where we have multiple
measurements we find that this scatter is typically of the order of 0.05 to 0.4 magnitudes. Only for the Mira variables we expect bolometric variations of up to 1 magnitude.

One star, RV\,Sgr, is clearly below the third dredge up limit and still shows Tc lines in its spectrum.
The K magnitude of this object has been taken from Whitelock et al.\,(\cite{wmf00}), who give an almost
complete near infrared light curve of RV\,Sgr. Therefore the difference cannot be explained by the
variability of the star. The Tc classification is from Little et al.\,(\cite{llmb87}). The star is an
oxygen rich Mira variable with a period of 317 days. In this context it is interesting to compare
the Hipparcos distances with those derived from photometric calibrations assuming that all Miras belong
to the same luminosity class (Celis\,\cite{celis95}). We found 9 M-stars stars among the AGB variables with good
Hipparcos parallaxes (according to our criteria) that are also found in the list of Celis. The photometrically
determined distances fall in five cases within the error bars given by the Hipparcos catalogue, in two
further cases the distances from Celis are only slightly outside the Hipparcos error bars. The remaining two
stars, o\,Cet and RV\,Sgr, show a somewhat larger difference between the two approaches, the latter one having the largest
offset within this small sample. The distance from Celis would shift the star upwards in Fig.\,\ref{f:hrd}
by 2.5 magnitudes, i.e.~clearly above the luminosity limit for the 3DUP\footnote{o\,Cet would be shifted
a little bit towards lower luminosities using the distance from Celis, but the star would still remain
in the 3DUP region.}. This higher luminosity is supported by the value expected from the period-luminosity relation
of Miras (Whitelock \cite{whitelock86}), which is almost identical. In the light of this
result we doubt the Hipparcos distance to RV\,Sgr. However, a more detailed analysis both of the distance
and the Tc occurrence for RV\,Sgr is mandatory to draw a definite conclusion on this star.

Another interesting result from Fig.\,\ref{f:hrd} is the fact that there are a number of stars clearly above
the 3DUP luminosity limit showing no Tc lines in their spectra. This result is independent of the 
variability class attributed. We will come back to this point later.

A further remark should be given on Fig.\,\ref{f:hrd}. It can be seen that Technetium is detected in stars
of all $J-K$ colors. This means that there is no dependency of the detection of Tc on the temperature of
the star. We can therefore exclude that any non detection of Tc lines may be due to a temperature effect.

\subsection{Period dependency of the 3DUP}
\begin{figure}
   \centering
   \includegraphics{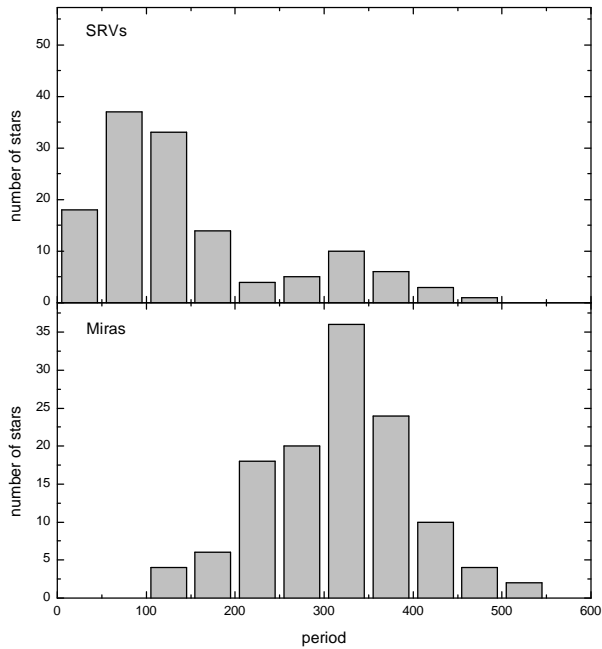}
   \caption{Histogram showing the period distribution of the SRVs (top panel) and the Miras
   (bottom panel) in our sample.}
   \label{f:perioddist}
    \end{figure}

A much larger sample could be used to investigate the dependency of the occurrence of Tc on the
characteristics of the variability, especially the variability period. As mentioned above
Little et al.~(\cite{llmb87}) concluded from their measurements that all Miras with periods longer than
about 300 days show Tc lines in their spectra. Lebzelter \& Hron (\cite{lh99}) found a few SRVs with Tc
at periods around 150 days. Data on long period objects, both Miras and SRVs, were rare. As a part of the
present project we could extend the list of long period objects significantly.
Figure\,\ref{f:perioddist} shows the period distribution of the SRVs and Miras in our sample.

    \begin{figure}
   \centering
   \includegraphics{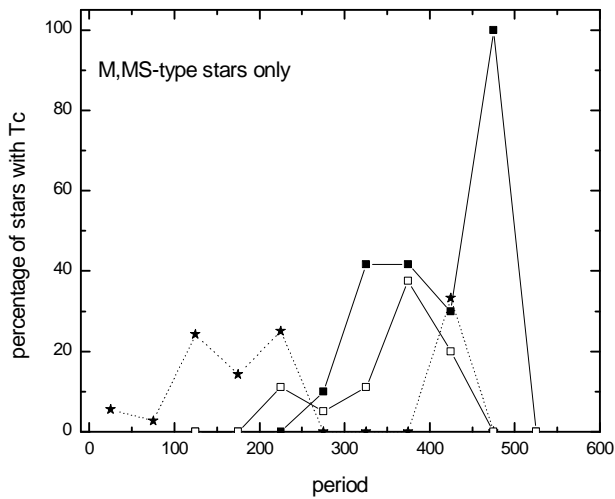}
   \caption{Fraction of stars with Tc lines within each period bin. Filled and open boxes mark Mira 
   variables clearly and possibly showing Tc, respectively (connected by a solid line). Asterisks
   mark semiregular variables (connected by a dotted line). Only stars with spectral type M and MS
   have been plotted here. For S- and C-type stars see text.}
   \label{f:fractiontc}
    \end{figure}
We calculated the fraction of Tc rich stars within each period bin separated by variability
type. The result is presented in Fig.\,\ref{f:fractiontc}.
It can be seen from Fig.\,\ref{f:fractiontc} that, opposite to the findings of Little et al.\,(\cite{llmb87}),
not all Miras with periods above 300 days have Tc. Taking into account only the clear cases, only about
45\,\% of the Miras with periods between 300 and 400 days show this signature of the third dredge up.
Even combining clear and possible cases, 100\,\% is never reached. For periods between 400 and 450
days the fraction of Tc rich stars drops even further. Only the four Miras with periods between 450 and 500
days all show Tc. Two Miras with even larger periods show no Tc. Semiregular variables with Tc are found
at periods between 100 and 250 days and around 425 days. The fraction of SRVs with Tc is typically of the order
of 20 to 30\,\%.

A statistical analysis (t-test) of the period distribution of Miras shows, as expected, a highly significant
difference between the group of Miras showing Tc and those showing not Tc. This reflects the fact that
Technetium is not found in short period Miras, which are thought to be metal poor (e.g.\,Hron \cite{hron91}).
On the other hand, the difference between Tc rich and Tc poor SRVs is not significant, i.e.~both groups show the
same period distribution. However, one has to take into account the small number of SRVs with Tc. In our sample
there are only 20 SRVs (13\,\%) with Tc.

%In Figs.\,\ref{f:lumiper1} and \ref{f:lumiper2} period and luminosity information are combined. No trends
%could be found. The low number of Miras in Fig.\,\ref{f:lumiper2} {\bf and the larger uncertainties in the luminosities due to parallax errors and variability for these stars do} not allow to draw any firm conclusions
%from this period-luminosity diagram.

%

%   \begin{figure}
%   \centering
%   \includegraphics{lumiper2.eps}
%   \caption{Period-luminosity diagram for Miras. Same symbols as in Fig.\,\ref{f:hrd}.}
%   \label{f:lumiper2}
%    \end{figure}
%
\section{Discussion}
We could show that the available observations are consistent with the luminosity limit for 3DUP as estimated from recent
stellar evolution models: only stars brighter than this limit show Tc lines
(excluding RV\,Sgr). However, our investigation also demonstrates that being brighter than this approximate
3DUP limit is not a sufficient criterion for a star to show Tc lines. This result is independent
of variability class although the fraction of stars above the 3DUP limit with Tc is dependent on
the variability type: Among the SRVs, only 15\,\% of the stars show Tc, while 44\,\% of the Miras
are Tc-rich. Irregular variables show almost the same fraction as SRVs.
We therefore have to ask, why obviously no Tc is detected 
in more than half of the stars above the 3DUP luminosity limit. For this analysis we will concentrate on the SRVs. 
The Miras have significantly larger luminosity uncertainties due to larger luminosity variations and parallax errors.

\subsection{Observational error sources}
We first want to discuss the different ingredients of our investigation.
Hipparcos parallaxes are currently the best known source for distance determination of individual objects.
Platais et al.\,(\cite{platais03}) have investigated the impact of wrong $V-I$ colors used in
the reduction of the Hipparcos measurements. However, for most stars there was no significant modification in the
astrometry when using correct $V-I$ values. Platais et al.~conclude that the effect is either insignificant
or it has somehow been accounted for. We have shown here that brightness variations due to stellar
pulsation provides only a minor effect.

For M and S type stars, high resolution spectroscopy allows to prove the presence or
absence of Tc lines in most cases. Possible problems in C-rich stars have been mentioned before.
In the present paper we found that the detection of Technetium is also obviously not dependent on the effective 
temperature of the star. 
We (Lebzelter \& Hron \cite{lh99}) have shown that also the measurements of Little et al.~(\cite{llmb87}) that
have been obtained at a somewhat lower resolution are -- with some exceptions -- reliable sources for the 
occurrence of Tc in red giant stars.
We therefore conclude that our approach is correct.

\subsection{Dredge up termination and metallicity effects}
The detection of Tc in the atmosphere of a red giant is an undeniable evidence of an ongoing s-process
within the star and a recent 3DUP event. On the other hand, there are several possible reasons for the absence of Tc. 
%is a clear indication
%that the star has not yet experienced a third dredge up (see the discussion in e.g.~Lebzelter %\& Hron
%\cite{lh99}). 
As can be seen from the calculations of Busso et al.~(\cite{bus92}) and Goriely \& Mowlavi (\cite{gm00}) the reduction of Tc during the interpulse
phase due to $\beta$-decay never endangers its survival. However, if no 3DUP takes place over several thermal pulses, the Tc produced earlier may decay below the detection limit (Busso et al.~\cite{bus92}). Such a situation could possibly occur if the envelope mass is reduced below the minimum mass for 3DUP. The solar metallicity 1.5\,M$_{\sun}$ models of Herwig et al.~(\cite{hbd2000}) may be an example for this. The termination of 3DUP could contribute to the apparent decrease of Tc-rich Miras with periods longer than 400 days. These stars are the most luminous, i.e.\ evolved objects among the Miras 
and mass loss could have reduced their envelope mass significantly.

Another possible effect preventing 3DUP is the influence
of metallicity. As discussed earlier, a lower metallicity increases the chance for 3DUP. If this behavior can be extrapolated to metallicities higher than the solar metallicity it could explain the lack of Tc in the stars above the 3DUP luminosity limit. According to the scale heights of SRVs and Miras (Kerschbaum \& Hron (\cite{kh92}), Miras (except for the short period
Miras) and SRVs should have a similar (solarlike) metallicity. Furthermore, the strengths of the metallic lines in the spectra of stars with and without Tc are very similar, indicating no large differences in metallicity.

\subsection{Influence of the stellar mass}
In Fig.\,\ref{f:hrd} we marked not only the approximate lower luminosity
limit for the third dredge up, but also the expected maximum luminosity during the thermal pulse cycle when 3DUP sets in. This maximum luminosity is the luminosity during quiescent hydrogen burning and should represent the stellar luminosity for one third to one half of the pulse cycle time (e.g.\ Straniero et al.~\cite{scl97}). Since on average the luminosity increases with core mass, all stars with core masses smaller than the limit for 3DUP discussed in Sec.~3.1 are expected to be found at luminosities smaller than this maximum luminosity. Stars with larger core masses should have luminosities above the luminosity minimum -~but not necessarily 3DUP because they may still have a too low envelope mass. Our stars with Technetium thus would have core and envelope masses above the limiting values, i.e.\ total masses of more than 
1\,M$_{\sun}$. 

For the stars without Tc which are brighter than the minimum luminosity there are two plausible explanations in terms of mass: (a) the stars have core (and maybe also envelope) masses smaller than the 3DUP-limit but are in a phase of the TP cycle with luminosities higher than the minimum surface luminosity (e.g. the quiescent hydrogen burning phase); (b) the stars have core masses higher than the 3DUP-limit (hence higher luminosities) but envelope masses below the minimum value required for 3DUP. Case~(a) would correspond to stars with a total mass below about 1.5\,M$_{\sun}$\footnote{Due to the lack of published models between 1 and 1.5\,M$_{\sun}$ more accurate constraints are difficult to set from this point of view.} in early stages of the TP-AGB. For case~(b) a wider mass range is allowed (i.e.\ also masses above 1.5\,M$_{\sun}$) but the stars have to be in advanced stages of the TP-AGB and the more massive objects would need to have high mass loss. Given the typical mass loss properties of SRVs (Olofsson et al.~\cite{olof02}), masses above 
2\,M$_{\sun}$ seem to be unlikely for the Tc-poor stars.

Combining all this we can therefore interpret our results in the sense that our sample contains a large fraction of stars with a current mass below 
1.5\,$M_{\sun}$. Alvarez \& Mennessier (\cite{am97}) have interpreted the scatter
in their temperature-period relation for Miras as a scatter in stellar mass. They
conclude that Miras show a mass range between 0.8 and 2.6\,$M_{\sun}$ with a mean mass of 1.4\,$M_{\sun}$. Since there are some indications that there is an evolutionary path from SRVs to Miras (see e.g.~the discussion in Lebzelter
\& Hron \cite{lh99}) this mass range is consistent with our results.

\begin{figure}
   \centering
   \includegraphics{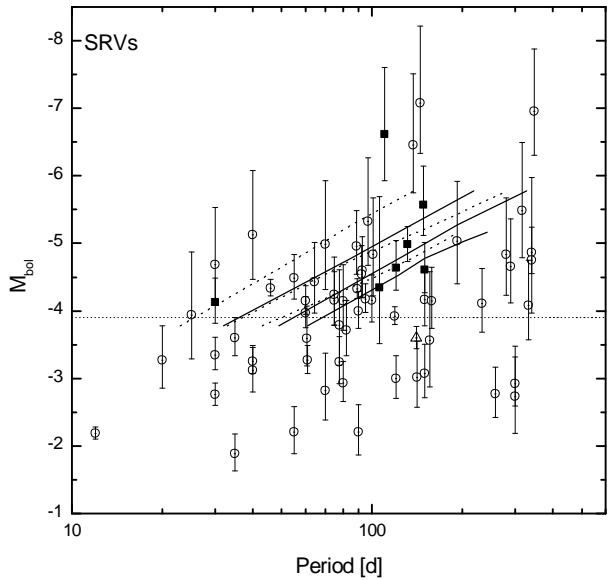}
   \caption{Period-luminosity diagram for SRVs. Same symbols as in Fig.\,\ref{f:hrd}.
   Solid and dashed lines mark the first and second overtone PL relations from Fox \& Wood (\cite{fw82}) for
   1.0, 1.5 and 3\,$M_{\sun}$, respectively (from right to left).}
   \label{f:lumiper1}
    \end{figure}

Independent information about the mass range of the SRVs may be deduced from a period/luminosity diagram (Fig.\,\ref{f:lumiper1}). In this diagram we included the theoretical PL relations for first and second overtone pulsations from the models of Fox \& Wood (\cite{fw82}). However, one has to remember that these models correspond to the quiescent H-burning stage, i.e.\ their luminosity is the maximum value attained during a TP-cycle for a given total mass. Thus when comparing the models with the observations one has to take that into account in addition to the parallax errors. Most of the Tc-rich stars are compatible with first overtone pulsation and pulsation masses of
1 to 2\,$M_{\sun}$. Two of these stars should have a higher mass and a higher pulsation mode. For the stars without Tc a wider range in mass and pulsation mode is required. While this fits with stellar evolution and nucleosynthesis for the stars on the right side of the 1\,$M_{\sun}$ first overtone PL relation, the lack of Tc-rich stars to the left of the 1.5\,$M_{\sun}$ first overtone PL relation points to pulsation of lower mass stars in higher overtones.  The existence of (at least) two pulsation modes among SRVs is also in agreement with findings from the LMC presented by Wood (\cite{wood00}) and from field SRVs by Kerschbaum \& Olofsson (\cite{ko98}). The single Tc-rich star at a period of about 30 days is $o^1$Ori. Although it's location would indicate a high mass it is a peculiar object (see the discussion in Lebzelter \& Hron \cite{lh99}).

The SRVs at bolometric luminosities above $-6^{\rm m}$ pose a problem since such luminosities would require masses of at least 3\,M$_{\sun}$, both from stellar evolution and from pulsation theory. For such masses the current stellar evolution models predict continuous 3DUP but only one star shows Tc. One of the Tc-poor objects (RV~Boo) is known to have peculiar mass loss (Bergmann et al.~\cite{bko00}). The other two stars (SV Peg, T Mic) seem to be quite typical SRVs. From their spectra and their galactic latitude a supergiant nature can be excluded. Their high luminosity may be due to an incorrect parallax but a further analysis of these stars is needed.

\subsection{Semiregulars and Miras}
Our results also have some implications on the variable stars on the AGB. Miras and SRVs occupy the same region 
above the 3DUP limit in Fig.\,\ref{f:hrd}. Still both groups have different fractions of Tc rich objects (about 
50\,\% of the Miras versus approximately 25\,\% of the SRVs).
A comparison of the occurrence of Tc and pulsational parameters like period or visual light amplitude
shows no indication of a correlation. This reduces the possibility that stellar pulsation (and the related
changes in the atmospheric structure) has some influence on the efficiency of the 3DUP.
It seems therefore likely that the difference between the two
groups reflects a different distribution of stellar masses. The average mass of the Miras would have to be somewhat higher than the one of the SRVs.
It should be stressed that the relevant parameter is the mass at the time when 3DUP sets in. This is therefore not necessarily the main sequence mass. 
The observed difference in Tc contents between the Miras and the SRVs may also indicate a different mass loss history, especially during the first 
giant branch stage.

There are some indications that there is an evolutionary path from 
SRVs to Miras (see e.g.~the discussion in Lebzelter
\& Hron \cite{lh99}). A combination of this scenario with our results suggests the following: The duration of the Mira and 
SRV stage -- relative to the total time on the TP-AGB -- is dependent on the stellar mass, i.e.~a star of 
smaller mass stays for a longer time in the SRV stage 
than a more massive star. One may suspect that below a certain mass limit a star never becomes a Mira. 

Further conclusions on this question cannot be drawn based on the data material currently available. Independent
mass determinations for a number of SRVs and Miras would be helpful. It also has to be stressed that statistical
data for the Miras are still based on a rather small sample of objects due to the lack of reliable parallaxes for
a larger sample.

\section{Summary \& Outlook}
We have shown that the occurrence of Tc in the atmosphere of an AGB star is not only dependent
on its luminosity being brighter than some 3DUP limit. Stellar evolution models suggest that the important second parameter is the envelope mass. The occurrence of Tc in AGB
stars can be understood in the view of some range in current stellar mass. The result could be checked by comparison
with constraints from stellar pulsation theory. A large fraction
of the stars therefore should have a current total mass of less than 1.5\,$M_{\sun}$. We suspect that
the absence of Tc in a significant fraction of long period Miras is due to a reduction of the
envelope mass below the critical limit by mass loss. To set further constraints on the mass
range of AGB stars stellar evolution models between 1 and 1.5\,$M_{\sun}$ would be needed.

We find that the lower fraction
of Tc rich SRVs compared to the Miras most likely reflects a lower mean mass of SRVs. 
Combining our findings with stellar evolution scenarios
we conclude that the fraction of time a star is observed as a SRV or a Mira is
dependent on its mass. 

In a forthcoming paper (Hron et al.~in preparation) we will compare our findings of
the present paper with a similar investigation on AGB stars in the Galactic Bulge.
With current instrumentation the Galactic Bulge represents the distance limit for
such an investigation. Currently planned instruments may be able to search for
Tc also in LMC Miras and beyond. This will then allow to test the 3DUP limit with
the help of Tc in environments of different metallicity.

Furthermore, one can hope that the small existing sample of Miras with reliable distances
can be extended by future space astrometry missions. 

\begin{acknowledgements}
TL has been supported by the Austrian Academy of Science (APART programme). Part of the data
used in this paper have been obtained at the AAO. We wish
to thank Stuart Ryder who did the service observations at the AAT. Furthermore, we are
indebted to Franz Kerschbaum for fruitful discussions and for providing his database of near infrared measurements and bolometric magnitudes. The constructive comments of the referee were very helpful for improving the paper.
We also acknowledge the support of Thomas Szeifert, Andrea Modigliani and Andreas Kaufer for
the preparation and reduction of the UVES observations.
\end{acknowledgements}


\begin{thebibliography}{}

  \bibitem[2002]{abia02} Abia, C., Dominguez, I., Gallino, R., et al. 2002,
      ApJ, 579, 817

  \bibitem[1997]{am97} Alvarez, R. \& Mennessier, M.O. 1997,
      A\&A, 317, 761

  \bibitem[1996]{bagnulo96} Bagnulo, S. 1996,
      PhD thesis, Queen's University Belfast, N. Ireland

  \bibitem[2000]{bko00} Bergman, P., Kerschbaum, F., Olofsson, H. 2000, 
      A\&A, 353, 257

  \bibitem[2002]{bkr02} Bergeat, J., Knapik, A. \& Rutily, B. 2002,
      A\&A, 390, 967

  \bibitem[1996]{bsw96} Bessell, M.S., Scholz, M. \& Wood, P.R. 1996,
        A\&A, 307, 481

  \bibitem[1992]{bus92} Busso, M., Lambert, D.L., Beglio, L., et al.\ 1992, ApJ, 399, 218

  \bibitem[1999]{bugawa99} Busso, M., Gallino, R., Wasserburg, G.J.\ 1999,
        ARA\&A, 37, 239

  \bibitem[2001]{bus01} Busso, M., Gallino, R., Lambert, D.L., Travaglio, C., Smith, V.V. 2001, ApJ, 557, 802

%  \bibitem[1979]{catchpole79} Catchpole, R.M., Robertson, B.S.C., Lloyd Evans, T.H.H., et al. 1979,
%      SAAO Circ., 1, 61

  \bibitem[1995]{celis95} Celis, L. 1995,
      ApJS, 98, 701

  \bibitem[1966]{cf96} Costa, E. \& Frogel, J.A. 1996,
      AJ, 112, 2607

  \bibitem[1997]{esa97} ESA 1997, 
      The Hipparcos and Tycho Catalogues, ESA SP-1200

  \bibitem[1982]{feast82} Feast, M.W., Robertson, B.S.C., Catchpole, R.M., et al. 1982,
      MNRAS, 201, 439

  \bibitem[1992]{fouque92} Fouque, P., Le Bertre, T., Epchtein, N., Guglielmo, F., Kerschbaum, F. 1982,
       A\&AS, 93, 151

  %\bibitem[1999]{gezari99} Gezari, D.Y., Schmitz, M., Mead, J.M. 1999,
  %     Catalog of Infrared Observations, 5th edition, available from Vizier at CDS

  \bibitem[1982]{fw82} Fox, M.W., Wood, P.R. 1982, ApJ 259, 198

  \bibitem[2000]{gm00} Goriely, S., Mowlavi, N. 2000,
       A\&A 362, 599

  \bibitem[1979]{hanson79} Hanson, R.B. 1979,
       MNRAS, 186, 875

  \bibitem[2000]{hbd2000} Herwig, F., Bl\"ocker, T., Driebe, T. 2000, Mem.\ SAI 71, 745

  \bibitem[2000]{HBSW00} Houdshelt, M.L., Bell, R.A., Sweigart, A.V., Wing, R.F. 2000,
       AJ, 119, 1424

  \bibitem[1991]{hron91} Hron, J. 1991,
       A\&A, 252, 583

  \bibitem[1995]{kerschbaum95} Kerschbaum, F. 1995,
       A\&AS, 113, 441

  \bibitem[1999]{kerschbaum99} Kerschbaum, F. 1999,
       A\&A, 351, 627

  \bibitem[1992]{kh92} Kerschbaum, F., Hron, J. 1992,
       A\&A, 263, 97

  \bibitem[1994]{kh94} Kerschbaum, F., Hron, J. 1994,
       A\&AS, 106, 397

  \bibitem[1996]{KH96} Kerschbaum, F., Hron, J. 1996,
       A\&A, 308, 489

  \bibitem[1998]{ko98} Kerschbaum, F., Olofsson, H. 1998, in M. Takeuti
  and DD. Sasselov (eds.), Pulsating Stars -- Recent Developments in 
  Theory and Observations, p.~177, Universal Academy Press, Tokyo

  \bibitem[1996]{klh96} Kerschbaum, F., Lazaro, C., Habison, P. 1996,
       A\&AS, 118, 379

  \bibitem[2001]{kll01} Kerschbaum, F., Lebzelter, T., Lazaro, C. 2001,
       A\&A, 375, 527

  \bibitem[1985--88]{GCVS} Kholopov, P.N., et al.~1985-88, 
	  General Catalogue of Variable Stars, 4th edition, Moscow (GCVS)

  \bibitem[2003]{KPPJ} Knapp, G.R., Pourbaix, D., Platais, I., Jorissen, A. 2003,
  				 A\&A, 403, 993

	 \bibitem[2002]{lattanzio02} Lattanzio, J.C. 2002,
	     New Astronomy Reviews, 46, 469

  \bibitem[1999]{lh99} Lebzelter, T. \& Hron, J. 1999,
      A\&A, 351, 533

  \bibitem[1987]{llmb87} Little, S.J., Little-Marenin, I.R. \& Hagen Bauer, W. 1987,
      AJ, 94, 981

  \bibitem[2003]{lug03} Lugaro, M., Herwig, F., Lattanzio, J.C., Gallino, R., Straniero, O. 2003, ApJ, 586, 1305

  \bibitem[1999a]{MGB99} Marigo, P., Girardi, L., Bressan, A. 1999,
      A\&A, 344, 123

  \bibitem[1999b]{mgwg99} Marigo, P., Girardi, L., Weiss, A., Groenewegen, M.A.T. 
      1999, A\&A, 351, 161

  \bibitem[2001]{mmal01} Mennessier, M.O., Mowlavi, N., Alvarez, R., Luri, X. 2001,
      A\&A, 374, 968

  \bibitem[1988]{mfop98} Montegriffo, P., Ferraro, F.R., Origlia, L. \& Fusi Pecci, F. 1998,
      MNRAS, 297, 872

  \bibitem[1999]{Mowlavi99} Mowlavi, N. 1999,
      A\&A, 344, 617

  \bibitem[2002]{olof02} Olofsson, H., Gonz{\'a}lez Delgado, D., Kerschbaum F., Sch{\"o}ier, F.L. 2003, A\&A, 391, 1053

  \bibitem[2003]{platais03} Platais, I., Pourbaix, D., Jorissen, A., et al. 2003,
      A\&A in press

  \bibitem[1988]{SL88} Smith, V.V., Lambert, D.L.~1988, 
      ApJ, 333, 219

  \bibitem[1990]{SL90} Smith, V.V., Lambert, D.L.~1990, 
      ApJS, 72, 387

  \bibitem[1997]{scl97} Straniero, O., Chieffi, A., Limongi, M.\ et al. 1997, ApJ 478, 332

  \bibitem[1998]{EJUMP98} van Eck, S., Jorissen, A., Udry, S., Mayor, M. \& Pernier, B. 1998,
      A\&A, 329, 971

  \bibitem[1986]{whitelock86} Whitelock, P. 1986,
      MNRAS, 219, 525

  \bibitem[2000]{wmf00} Whitelock, P., Marang, F. \& Feast, M. 2000,
      MNRAS, 319, 728

  \bibitem[2000]{wood00} Wood, P.R. 2000,
      PASA, 17, 18

  \bibitem[1996]{ws96} Wood, P.R., Sebo, K.M. 1996,
      MNRAS, 282, 958

\end{thebibliography}
\end{document}